# Increasing the Likelihood of Finding Public Transport Riders that Face Problems Through a Data-Driven approach

Vasco Furtado, Carlos Caminha, Elizabeth Furtado, André Lopes, Victor Dantas, Caio Ponte, Sofia Cavalcante

*Programa de Pós Graduação em Informática Aplicada, Universidade de Fortaleza*
*Av. Washington Soares, 1321 - Edson Queiroz, 60811-905, Fortaleza, Ceará, Brasil.*

**ABSTRACT**

The maintenance of big cities' public transport service quality requires constant monitoring, which may become an expensive and time-consuming practice. The perception of quality, from the users' point of view is an important aspect of quality monitoring. In this sense, we proposed a methodology for data analysis and visualization, supported by software, which allows for the structuring of estimates and assumptions of where and who seems to be having unsatisfactory experiences while making use of the public transportation in populous metropolitan areas. Moreover, it provides support in setting up a plan for on-site quality surveys. The proposed methodology increases the likelihood that, with the on-site visits, the interviewer finds users who suffer inconveniences, which influence their behavior. Simulation comparison and a small-scale pilot survey stand for the validity of the proposed method.

## 1. INTRODUCTION

As part of any metropolis' daily routine, many are the public transport (PT) riders who select a less than optimal path from their origins to their destinations. Literature points out that path choice, as part of transport demand modeling, is based on performance measures (such as time, distance, number of connections, etc.) [1] in terms of level of service [2], at the same time, it can be influenced by the user's perception of quality [3]. The users' perception (and the environmental characteristics that trigger them) are so important they can be related to demand elasticity measure [4] and in many cases are mirror to the attitudes of transport users [5], including PT riders. The systematization of these perceptions is often a result of qualitative surveys aiming the subjective opinions of users.

In the last ten years, many were the studies that focused on evaluating PT service quality [3,6–10]. The scientific literature has also an abundance of developed methods, techniques and tools to conduct field surveys to capture user preferences [11,12]. Challenges to perform this monitoring start with the fact that there is still no consensus on the main elements that make up this notion of perceived quality. The elements used to measure it can be both subjective (comfort, safety, etc.) and objective (availability, accessibility, etc), with 'service reliability' and 'point-to-point travel time' as two key elements influencing such decisions [13]. Recent studies have shown that the importance given by users to these factors varies depending on the city and the context [14]. This corroborates design theory that suggests elevated dependency levels between design and local contexts, the importance of design to human experiences and perception [15], and that attitude and behavioral patterns (specifically for transport modal choices) are correlated with user perception of environmental characteristics [5,16].

The importance of service quality evaluation rests on the fact that such information is necessary for planners, decision makers and operators to propose and run high performance PT services. This evaluation should take into consideration passengers' priorities, as well as the service quality, not only when related to level of service, but also in terms of riders' perception of quality. As an evidence of this importance, in recent years, many PT operators and decision makers adopted quality-based incentives so that expected quality (from rider's perception) and real quality (presented by the system) would converge. One difficulty to achieve such system melioration is that data collection on subjective perception of riders is usually collected through surveys, whereas objective performance measurements are automated [17]. While the amount of data automatically collected can easily scale up to millions of observations, quality surveys are slower to collect and more costly.

In this context, we pose three questions. First, is there a more objective way to identify PT problems related to the riders' perception of system's quality? Second, what kind of data we need and how can we process it so that we may easily identify rider's perception-related problems? Third, can we turn satisfaction surveys into a problem-driven tool that is more efficient and less costly? To answer such questions, we adopted a data-driven approach, taking advantage of a heavily monitored PT system in a 2,5 million inhabitant metropolis that counts with a 300,000 bus trips per weekday, all of which are GPS tracked every 30s, and from which more than 1,7 million individual origin/destination pairs could be identified a day. This amount of information, once submitted to the proposed method of analysis, allowed us to pinpoint potential unsatisfactory experiences from PT riders in a more precisely way if compared to regular data-collection methods. These results allow us to implement more efficiently quality-related surveys, from which we may identify drawback PT services imposed over riders.

We organized the present paper in six sections, already counting the introductory first section (1). In section two (2), we conduct an analysis of related works, focusing on papers that integrate data analysis and preference surveys. From this overview of literature, in section three (3), we formalize the PT riders' path-choice problem and propose a heuristics so that we may identify and quantify discrepancies between the systems' performance (in terms of potential optimal routes) and PT riders' quality perception (in terms of attitudinal decisions). This is followed, in section four (4), by an empirical application of the method, where we describe the available data, quantification of PT riders' choices, and the tools to represent resultant discrepancies when compared to optimized path-choices. The fifth section (5) brings some evidence to validate the method regarding its effectiveness in finding those users regarded to experience unsatisfactory PT services (those that choose less than optimal routes). In the sixth and final section (6), we discuss the repercussions of such results in terms of quality perception data gathering, and more cost effective survey methods.

## 2. DATA ASSESSMENT FOR 'PT' QUALITY SURVEY

This section makes a comparative analysis between studies that deal with objective and subjective data to assess the quality of public transport services. The comparison focus on the support provided for a qualitative and quantitative data exploration. We understand this exploration as a cyclic process, which comprises the tasks of data collection, processing and analysis, and hypothesis testing. As the solutions evolve and a large data sample is collected, the qualitative and quantitative approaches that follow this cycle tend to converge, from majoritarian focus on user-centered analysis to a more balanced data/user-driven approach. Quantitative data helps researchers to identify patterns of human behavior (including PT riders), while qualitative data (e.g. from interviews with participants) allow us reveal their attitudes and perceptions. Corroborating Wright (2015), we understand that quantitative data is a powerful tool that can help researchers recognize spatial patterns, and direct them to specific areas where an optimized survey can be carried out in order to talk to people who are able to reveal specific information previously unknown, or "hidden" in the data. We could envision, at least, three methods that favor a user-centered investigation directed at quantitative data. Researchers could make regular use of locating trackers to collect data where people (research subjects) access the service or product under investigation [19]; they could combine data-mining methods and visualization techniques (infographics, dashboards) as support tools for decision-makers. The dynamic aspect of these different areas (knowledge discovery, visualization, and decision support) creates uncertainties that need to be tested, not limiting the analysis to the decision-makers subjective judgment [18]; and finally, they could use contextualized issues. When the user's opinion is important, assumptions about users' preferences help prepare the questions that focus on the problem to be investigated [20]. Problems may change according to the location and time of investigation.

For the sake of comparison, we divided the tools proposed in each of these methods in two groups. First, the supporting tools for (objective and subjective) data collection, analysis and integration, wherever possible, and second, the supporting tools for contextualized analysis of public transportation. The dynamic nature of PT supply and complexity of human behavior require the definition of what criteria should be used to define quality. Furthermore, we subdivided each of these two groups in three subgroups, following the tasks of (1) data collection, (2) processing and analysis, and (3) hypothesis testing.

The first observation is that none of the analyzed studies provides support tools for any qualitative and quantitative cyclical analysis. When a tool is employed, it is an isolated support for specific tasks. For the data collection task, it is important to emphasize that while current technology makes surveys increasingly available in many places, only two studies [11,12] made use of web technology as a supporting tool. In other studies, researchers directly applied surveys to obtain user satisfaction compared to the quality criteria. Bajaj et al. (2016) made use of a mobile application to collect the preferences of PT users. In the Brazilian context, much of the captive PT riders have no access to sophisticated mobile phone technology (like GPS or Bluetooth), moreover, privacy issues may also pose as an additional difficulty for implementing such tools.

For the processing and analysis of quantitative data, there is a tendency to use learning and mining techniques aimed at recommending routes for users to follow while using PT services. Recommendation systems [19] and/or systems to guide users (like Google maps), are outside the context of this research. The study by [21] was considered because it collects feedback from choosing a route recommended by the system, this is useful information to evaluate the quality of the user experience with the preferred route. No tool was identified that could assist in the investigation of user preferences specifically in the formulation of hypotheses that guided the generation of questions relevant to the local research.

In another analytical framework, we divided the analyzed papers into four categories of service-quality assessment, according to the framework of Nathanail (2008). They are, (a) Experienced quality (perceived); (b) Desired quality (expected); (c) Provided quality of service; and (d) Continuous improvement of service quality. Some authors [22] apply a methodology to record PT (potential and actual) users' desired quality. The desired quality is different from the experienced quality, as the first only means what users want or expect from the system, and the second should represent users' daily experiences, dealing with the users' emotion in a dynamic context [23]. Provided quality of services means the systems' actual measure, which could be related to what transport researchers would call the level of service. The fourth category should interest PT system' managerial and decision-making staff, as it is focused on continuous quality improvement. This framework served as classification criteria for contextualized supporting tools and methods concerning location and time for survey application. In most reviewed studies, the context used for the data collection has been previously defined [11,22,24,25], except for those where the user is guided by an application.

Setting the time and place for questionnaires/surveys application aiming the collection of users' opinion is still an open debate. In addition, questionnaires measure only preferences or intentions (attitudes); they do not say much about the way people behave [26]. As for the investigated quality criteria, recent studies have drawn attention to the extensibility of these criteria [13,21], while most authors adopt pre-defined criteria, regardless of location and time. Tyrinopoulos & Antoniou (2008) adopted a methodology that followed various strategies such as focus group, survey pilot and model calibration to determine and test the most relevant inconvenience variables before applying the survey. In addition to pre-defining the criteria used in an investigation, many studies also feature solutions to synthesize the research data. Furthermore, regarding the identification of users, in the methodology proposed by [22] there is an effort on the part of the researchers to reach an adequate sample according to geographic area. Most studies aim to overcome the difficulties in the construction of sampling frames and to ensure that respondents are not self-selected [26]. This is because traditionally the work is intended to draw from a population sample, with general characteristics of the population as a whole. Table 1 summarizes this whole discussion, presenting a side-by-side comparison of supporting tools.

Table 1: Synthesis of supporting tools and contextualized analysis for PT users' preferences

| **Supporting tools** | **Support for contextualized analysis** |
|---|---|
| *For the data collection task* | *Data collection sites, and surveyed users for service quality assessment* |
| Bajaj et al. (2016): GPS assisted mobile application (Metro Cognition) for users' feed-back about the convenience of recommended route;<br><br>Vovsha et al. (2014): Online survey tool to collect riders' evaluation about current route (punctuality, vehicles load, etc.) and generate alternative routes. Riders must justify the route choice decision. It also gathers email addresses, geocoded locations and starting travel times. Down side: extensive (+50 questions) survey.<br><br>Nathanail (2008): Online survey application to collect operational indicators (the transit system performance). | Focus on quality of experienced service: Tested and collected in real time by 24 volunteer users: The context depends on the route chosen by the user of an application, because it chooses a recommended route, and gives feedback on the traveled route [21].<br><br>Focus on quality of desired service: For 305 different users profiles and 200 potential users [22]. The collection is made in a pre-defined context (the buses and the stops of Santander and at peak hours on working days and on streets).<br><br>In [12], 2069 Los Angeles metro users answer a survey about their preferences.<br><br>Focus on quality of provided service: in [25], a sample of 123 users of a bus line that connects two cities in Italy, they give satisfaction marks on their perception of the service.<br><br>[11] also defined sample size and where to apply the survey (1471 users found in 5 transit systems in two cities in Greece).<br><br>Focus on continuous improvement of service quality: In [24], 13 employees of a company expressed their feelings about being and to continue being users of public/shared transport. |
| *For data analysis and visualization task* | *Criteria used for data analysis* |
| Bajaj et al. (2015 and 2016): Use learning and mining techniques for route recommendation, based on PT riders' preferred convenience criteria;<br><br>DellOlio et al. (2011) and Vovsha et al. (2014): Tool for statistical analysis of criteria importance, categorized by type of PT rider (age, gender, etc.). Tyrinopoulos & Antoniou (2008): did the same for satisfaction levels; | Service (in)convenience criteria are predefined [11,12,14,22,25,27] and place and time independent;<br><br>Preferences of captive PT users, in the form of users opinions, collected while the service is in use (Castellani, 2016);<br><br>Eboli & Mazzulla (2011) explore the relationship between user satisfaction and service quality criteria (pre-defined in [11]).<br><br>Bajaj et al. (2016) and Nathanail (2008) analyze service quality based on users' perceptions of criteria that guide their choices of route alternatives. Eboli and Mazulla (2011) do the same, but include the perception PT systems managers and operators; |
| Nathanail (2008): Tool for data storage and calculation of indicators for convenience criteria; | |
| *For synthesis and hypothesis formulation task* | *Synthesis and hypothesis tasks that guide quality assessment* |
| Bajaj et al. (2015 and 2016): Real time monitoring of users' habits. Data visualization for managerial purposes depends on application connectivity during its use.<br><br>Castellani et al. (2016): Seeks a more personalized service, allowing filtering users' certain behaviors. Tracking the users' travel history allow the development of hypotheses about the changing habits of users (still under construction). | For the routes recommended by the researcher: The researcher predicts the level of convenience for the recommended routes, which are evaluated by feedback from users (Bajaj et al. 2015 and 2016);<br><br>For scenarios described by users. Scenarios are used to define the criteria to be investigated, thus simplifying the collection of user preferences [22];<br><br>For the location of users. In [12] and [14], the automatic user location helps synthesize what to ask; and<br><br>The correlation between attributes of convenience criteria. [11,25] estimated 22 attributes of service quality from users perception of quality and managers (quality measurement), before applying a survey. They evaluate the correlation between attributes to simplify the analysis of user satisfaction when compared to the relevant criteria. |

This overview reaffirms that the combination of qualitative and quantitative approaches supported by data-mining technologies that help in can be a very helpful approach. The selection of users who potentially had an unsatisfactory experience is an inexpensive, objective and effective way to find problems in the public service, as well as locate inconveniences previously unnoticed.

## 3. DESCRIPTION OF PROBLEM AND PROPOSED SURVEY METHOD

Route choice modeling stands as a common challenge for transport planners. The description of a route choice behavior entails the modeling of individual route decisions from sets of alternative feasible routes. In general, the number of alternative routes is rather large, but, when dealing with PT ridership, the identifiable alternative routes between each OD pair are restricted to predetermined lines' itineraries and transfer points. Defining the feasible routes to form a set of alternatives that satisfy an OD pair is understood as a preference-driven and constraint-related choice, since undesirable characteristics may lead to route elimination [28]. Differently, choosing a route from a set of alternatives is a compensatory action (as a trade-off between characteristics of different routes), normally leading towards an optimization problem. Nonetheless, chosen routes may or may not satisfy PT rider's necessities and/or desires. With this conceptualization in mind, we propose a formalization for our route-choice problem.

### 3.1. Problem Formalization

We see the transport of a user 'u' from origin 'o' to destination 'd' by a vehicle that follows a predefined route 'r' as function, f : R x P, where 'R' is the set of alternative routes known to the user to take him/her from 'o' to 'd', and P is the set of preferences attribute to the user 'u'. The preferences 'p' are based on criteria such as travel time, number of connections, etc. Thus, there are users who prefer routes that lead them from 'o' to 'd' in the shortest time, while others may prefer to reach the same destination by a route where there are the fewest connections. Formally, each user 'u' from their respective origins 'o', choose a route '$r_i$', so that R = \{$r_1$, $r_2$, ..., $r_n$\} based on criteria to measure the convenience of the service, 'k' of each OD pair, where k ∈ K=\{distance, time, number of connections, number of stops\}. The function $\text{valc}_k(r_i)$ gives a value for the criterion 'k' of the route '$r_i$'. For example, if k = 1 and the distance traveled by a vehicle to make the route, '$r_1$' from 'o' to 'd' is 20km, so $\text{valc}_1(r_1)$ is 20.

For each choice made by a user from the origin, it is assumed that 'u' has a preferable relation Pref of $K_p \in K$ to all other $K_t$, $t \neq p$, $K_t \in K-\{K_p\}$. This relationship indicates the preference of a criterion in relation to all the other denoted $K_p$ Pref $K_t$. Consequently, there is a relation of preference between the values of the preferential criterion 'p', for each route, $V = \{valc_p(r_1), valc_p(r_2), valc_p(r_3), ..., valc_p(r_n)\}$, which is denoted by $valc_p(r_i)$ Pref $valc_p(r_j)$, for every $j \in \{1, 2, ..., n-1\}$ and $r_j \in R - \{r_1\}$. Assuming that $valc_1(r_1)=20$ and that there is another route, '$r_2$', with $valc_1(r_2)=30$ and another, '$r_3$', with $valc_1(r_3) = 40$, then a user who wants to take a route minimizing the distance traveled will consider that the route '$r_1$' is preferable to routes '$r_2$' and '$r_3$', or, formally, $valc_1(r_1)$ Pref $valc_1(r_2)$ Pref $valc_1(r_3)$. Once we have the alternative routes, related to the constraining preferences of users, we may discuss the choosing of a route from the alternative set. The literature [2,28] points to the common practice of applying optimization algorithms on different types of selection criteria as a way to model such chosen routes. For this experiment, we have several data sets that describe the chosen routes for each OD pair. One set from empirically collected data, named as real routes (rr), and several other sets resultant from optimization processes that took into consideration different criteria, named as optimized routes (ro).

Finally, we should define if chosen routes do or do not satisfy user's expectations. Considering a satisfaction threshold applied to the routes' convenience measure (such as distance, travel time, number of connections, etc.) it is possible for the analyst or planner to recognize if PT riders having a unsatisfactory experience (E.g. assuming a 10min threshold, riders who prefer less time-consuming trips but choose a 10min (or more) longer trip might feel unsatisfied). It is worth mentioning that until this moment, we cannot state with certainty that the rider actually had an unsatisfactory experience. This classification only shows that, within a population that values travel time there are some riders choosing routes that lead to at least 10min longer trips. We believe these are potentially good subjects to be surveyed, as they may have been experiencing unforeseen inconveniences, not accounted by the applied route choice model, which could explain their behavior. The proposed heuristics takes into account a threshold of satisfaction ($\lambda$) defined by the analyst, which is dependent on context, culture, demographics, etc. For each user's real route (rr) the value of a particular evaluation criterion ($valc_i(rr)$) will be subtracted by the value of the same criteria attributed to the optimized route ($valc_i(ro)$). Thus, the classification of the experience C, from a user's actual route (rr) could be satisfactory or unsatisfactory and may be defined as:

$$C = \begin{cases} \text{satisfactory}, & valc_i(rr) - valc_i(ro) < \lambda \\ \text{unsatisfactory}, & valc_i(rr) - valc_i(ro) \geq \lambda \end{cases} \tag{1}$$

**3.2. Proposition of Survey Method**

To try to answer the three questions posed in this paper' introduction, we developed a four-step cyclic methodology. It poses as useful tool to help planners and decision makers to fulfil the tasks of identifying PT ridership related problems, of deciding where to gather necessary data, and defining how the perception of the environment may influence PT riders' behavior. Figure 1 brings a representation of this four-step methodology.

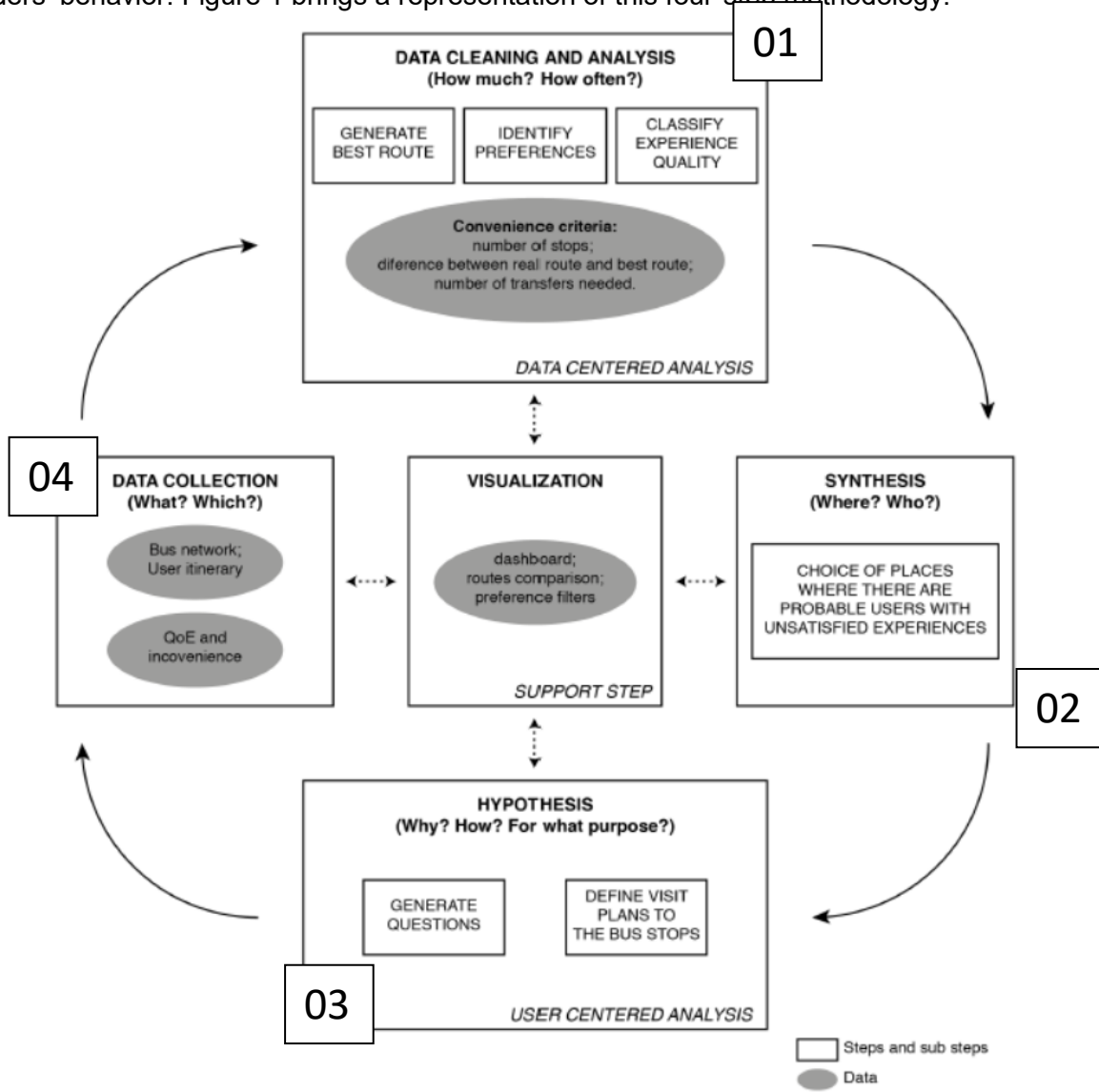

FIGURE 1. Steps of Proposed Survey Methodology

1. The first step refers to the data cleaning and analysis. The data comes from PT card validation, made by each passenger at the vehicles' built-in turnstile and terminals, and is processed through data mining processes combined with visualization techniques. This step consists in three sub-steps of (a) optimal routes generation; (b) riders' preferences identification, and (c) the classification of riders' experiences.
2. The second step aims to support the synthesis of the research in regards to where to investigate and who are the potential users with unexpected behavior;
3. The third step initiates a more user-centered analysis approach. At this point, the researcher must establish hypothesis about what drives PT riders in choosing a routes;
4. The fourth step consists of on-site data collection, based on questionnaires applied at preselected bus stops, with "in route" users (at the time that they make route choices). The researcher should aim to contextualize a survey about riders experience in using PT. Riders may confirm and/or refute the assumptions made or clarify other disregarded aspects that would justify the posed hypothesis.

Moreover, we propose a supporting step that contributes to the monitoring of the process. It consists in a visualization tool, which decision-makers can use during all other steps of the method, and may constantly search for information and clarification on routes and PT riders' preference. Next, we advance on describing the steps related to the data centered analysis, comprising the recognition of optimized routes, riders' preferences, quality of experience and finally the location of potentially unsatisfied PT riders. It corresponds to steps 1 and 2 of the proposed method.

### 3.3. Sub-step (a): Generating Optimal Routes

To generate an optimal route for a PT system it is necessary to have enough information about its OD pairs. For this task, we rely on tracking PT riders' actual routes, as well as on data from the city's transport supply networks. In some cases, the tracking of PT riders is done in such a precise way that the initial sample can represent up to almost the entire population of passengers [29,30]. In most cases, however, such a sample needs to be estimated either by information of traffic [31], [32] ticket registration of passengers and/or an origin-destination matrix (OD) obtained from interviews with citizens. Furthermore, any route optimization requires at least one aspect of the route to optimize. Without limiting the generality, we consider here four optimization objectives, which, deliberately, coincide with the convenience measures: the shortest possible distance (1), shortest time (2), the lowest number of bus connections (3) (transfers), and take the route with fewest stops (4) possible. The methodology accepts other objectives, which depend on context. We calculate the optimal routes for each objective. Figure 2 presents the search algorithms of optimal routes as 'graphs' representing the available bus lines. The graphs' nodes are bus stops and the paths between stops are the arches. From many graphs, representing individual lines, one may construct a graph that would represent the whole system.

In figure 2a you can see the representation of a bus line *L1* that has in its itinerary four bus stops represented by the nodes, $v_3$, $v_4$, $v_5$ and $v_6$. In Figure 2b, the bus line line *L2* presents a different itinerary with the nodes $v_1$, $v_2$, $v_4$, $v_5$, $v_7$ and $v_8$ representing its stops. Finally, in Figure 2c, we have the two combined lines. Note that there are both arches and nodes that are shared between both lines L1 and L2, meaning an increase in its weight. The graph is completed once all the existing bus lines are overlaid on the graph, representing the transport supply network.

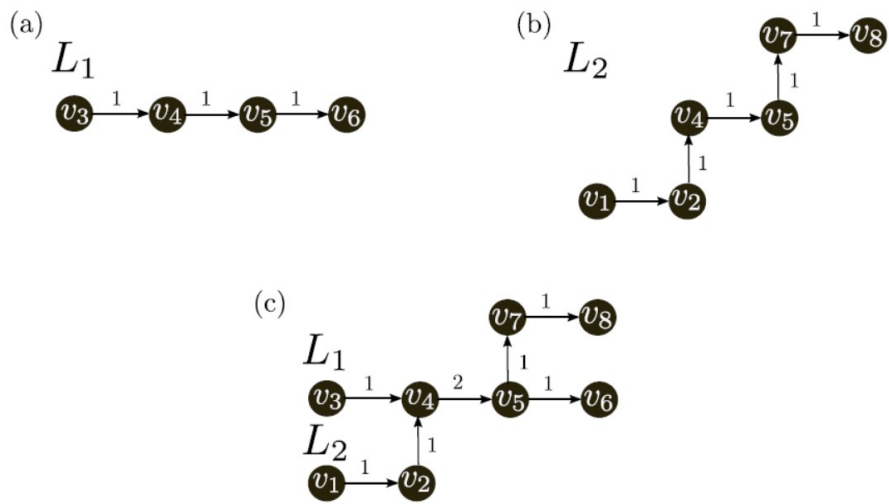

Figure 2: Example of a graph constructed from the stops and bus lines.

The optimization of routes for each of the four predetermined objectives demanded different algorithms. The calculation of optimal routes by distance may be done using traditional search algorithms on graphs as such A* [33]. Regarding the estimated route for users who want to use express lines, or lines that go through fewer bus stops, one can estimate optimal routes using a spread search [34]. Different from the first two objectives, the estimated optimal routes for fewest connections (transfers) and for the shortest possible time require a little more detail in the description. This is because there are no automated algorithms prepared for such tasks. We developed two algorithms able to find: i) the route between any two bus stops making the least amount of bus changes; and ii) the route between two points in the shortest time possible. Appendix A contains the description for these algorithms. It is very important to stress that we do not see these algorithms as a contribution of this paper. Researcher may use any optimization algorithms, as long as they apply the proposed survey methodology.

**3.4. Sub-step (b): Identification of User Preferences**

Preference surveys typically use statistical approaches, meaning that they rely on sampling the population under study. Although surveys are important sources of data collection, they can be very slow to gather, as well as to tab and analyze. We believe that, concerning route choice preferences, data gathered directly from the observation of user's behavior should provide information in more quantity and detail. Once we have reliable data on observed route choices (in the form of OD pairs), we propose an algorithm that points what convenience measure each PT rider have when choosing bus routes. It consists in comparing the actual routes taken by each observed rider (comprising its OD pair) and each of the generated optimal routes, fallowing the four objectives that were to minimize distance, travel time, vehicle changes, and hops. We assume that the more similar to one optimized route objective, the higher the chance that objective is the route choice criteria adopted by that user, indicating his/her preference.

Although we could use any method for calculating the difference between actual and optimized routes, we propose an algorithm based on the area comparison of polygons representing aspects of these routes, in the form of "Kiviat diagram". The polygons are formed by the join of edges, of which the vertices indicate the measurement of one of the four objectives used for generating optimized routes. Figure 3 illustrates a comparison of setting routes for an arbitrary OD pair, where we calculated the intersection of a polygon formed by the actual route with the optimal routes for each one of the four objective.

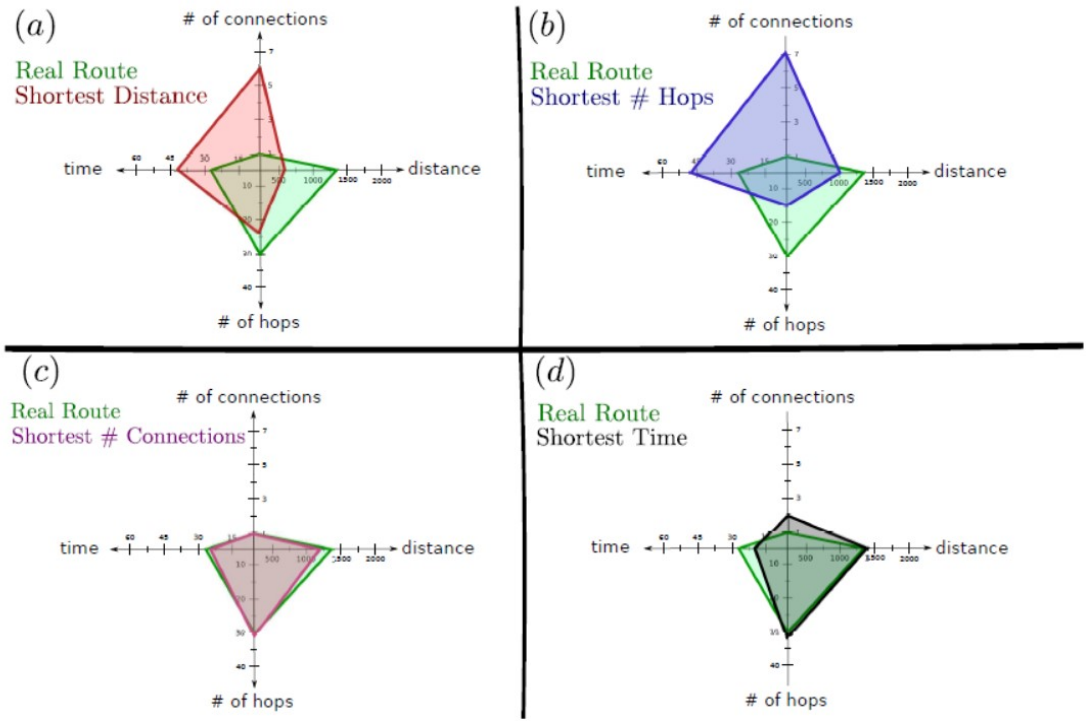

Figure 3: Polygons intersection on Kiviat diagrams. The actual route is compared with the optimal route regarding shortest distance (a), number of hops (b), number of bus transfers (c), and shortest travel time (d). The highest similarity level happens in (c), which contains the highest intersection between polygons, indicating the user's preference lays on minimizing the number of transfers.

### 3.5. Sub-Step (c): Quality Classification of Users' Satisfaction

Once we recognize the riders' preferences, among the predefined objectives, we may further compare how different actual and optimal routes are. This should allow us to verify whether the rider might be experiencing a unsatisfactory experience (E.g. a person whose preference is for trips with fewer transfers but actually takes a route with more transfers than the minimum is a potential unsatisfied PT rider). At this point, we may apply equation (1), presented in the problem formalization, to define whether the situation is satisfactory of not.

### 3.6. Step 2: Localizing unsatisfied users

Our idea to estimate which bus stops are more likely to have people in the condition of dissatisfaction is based on a probabilistic approach. We may formally define the probability *P(p)* of finding users who had unsatisfactory experiences at a stop *p* as:

$$P(p) = \frac{Qr_p}{Qr_p + Qb_p} \qquad (2)$$

leaving $p$ and $Qb_p$ is the quantity of people who took routes that led to a satisfactory experience at the same bus stop. Using the available visualization tools (in this case, we developed a dashboard for monitoring the process) the recognized outliers are plotted on a map of the PT network. This approach allows decision makers to visualize whether the occurrence of this type of behavior is spatially dependent to some other phenomenon, or specific to a region of the city. Graphs depicting the distribution of these potentially unsatisfied PT riders should help researchers in selecting where surveys should take place so that they have a higher probability of finding user with such profile.

## 4. APPLICATION OF THE METHOD

The methodology described above was applied in Fortaleza, a large Brazilian city, where about 1 million people are frequent PT users. For this research, we used five data sets (bus stops; lines; terminals; bus GPS, and ticket validation) related to the public transportation network of Fortaleza (Table 2). The bus GPS data has 104 million positions recorded from the location of the buses, which are updated every thirty seconds. Altogether 2,034 buses circulated on 359 bus lines in the city in March 2015. The tickets are validates by magnetic card readers. The system registered 29 million ticket validations in busses or directly at one of the seven existing terminals. Registered data consists in 'user registration', 'date' and 'validation time', 'bus-line or terminal number' and 'vehicle code'.

Table 2: Group of Used Data

| Name | Registered Numbers |
| --- | --- |
| Stops | 4783 |
| Bus Lines | 359 |
| Terminals | 7 |
| Bus GPS | 104 Millions |
| Tickets | 29 Millions |

For the studied city, we had to estimate the origins and destinations of the users as well as reconstruct the actual routes from a sample, from which we could identify the OD pairs. The estimates derived from data gathered from PT card validations that were transformed in OD pairs through a estimation method proposed by (Caminha, Furtado, Pinheiro & Silva 2016; Caminha, Furtado, Pinheiro & Ponte 2016; Caminha, Furtado, Pequeno, Ponte, Melo, Oliveira, et al. 2017) [35–38]. The raw data did not allow us to know precisely from which bus stop the passenger boarded the bus, it only contained the time that a particular user validated his ticket at the buses' built-in turnstile or terminals. From this data, we had first to estimate where this validation happened. The method, described in the referred literature, lays beyond the scope of the article. As a result, we obtained 1,737,772 OD pairs from the bus network for a week in March 2015, from which about 1.2M of actual routes were reconstructed.

### 4.1. Estimated preferences of users

With the actual routes at hand, we applied the polygon intersection (PI) approach for the routes comparison. For the sake of method assessment, we also conducted another comparison method, which is based on calculating the Euclidean distance (ED) between the properties of routes 9. Table 3 shows the results for both comparisons. The results are consistent with each other, providing evidence that the proposed method is consistent with other well-established methods, and that a bigger part of PT riders prefer to use routes with fewer transfers than any other optimization objective.

Table 3: Route Comparison Results

|  | PI | ED |
|---|---|---|
| Best transfers | 38.2% | 36.2% |
| Best in time | 20.8% | 21.2% |
| Best distance | 9.5% | 11.2% |
| Best hops | 2.1% | 3.9% |

Besides the numbers described in table 3, there were also similarities in the cases where the actual routes were identical to the estimated route by more than one criterion. This occurred in cases where the distance between the estimated origins and destinations were very small. In these cases, due to the network characteristics, algorithms tend to converge in the same direction.

### 4.2. Riders behavior patterns and locations selected for data collection

To help us analyze all collected data, we introduced an analytical dashboard as the supporting step of the proposed methodology. From a number of analysis enabled by the dashboard (such as, passenger volume, timetables, routes, distances, and user profile evaluation) we could recognize that a maximum of three transfers per route were made in the city. Another major discovery was that most riders would transfer at bus terminals, meaning that they do not choose the best place to transfer (standard increment behavior), when compared to the optimized routes, resulting in longer than necessary routes.

With the help of the supporting dashboard, we filtered selected discrepant routes with the satisfaction threshold ($\lambda$) (preference-criteria differences between actual routes and those generated by the optimal algorithm). Even though observed criteria for the estimated preferences of users pointed to the number of transfers as the best indicator, for this example, as a way to clarify the validation efforts, we measured $\lambda$ in terms of distance. It is important to highlight that researchers/decision-makers could adopted any indicator as preference criteria. We filtered routes, which distances differed more than 2km from one another. This analysis resulted in heat maps, identifying stops where the probability of finding people potentially unsatisfied with the PT ridership experience was greater. From this synthesis, we derived a hypothesis that these PT riders did not choose the best routes due to the lack of local information about possible routes. Next, we could put together a survey planning, defining to which stops researchers should apply the questioners and collect information.

We provided the interviewer with reports produced by the dashboard, indicating which stop he/she should go to, and was given details about the actual route, the optimal route and a detailed description of the steps that he/she should follow to complete the optimal route. In these detailed reports, there is information about origin and destination, the distances involved in each of the two routes (actual and optimized for the riders' preference), the number of transfers and the bus lines to be used in the presented suggestion. Figure 4 illustrates a full report.

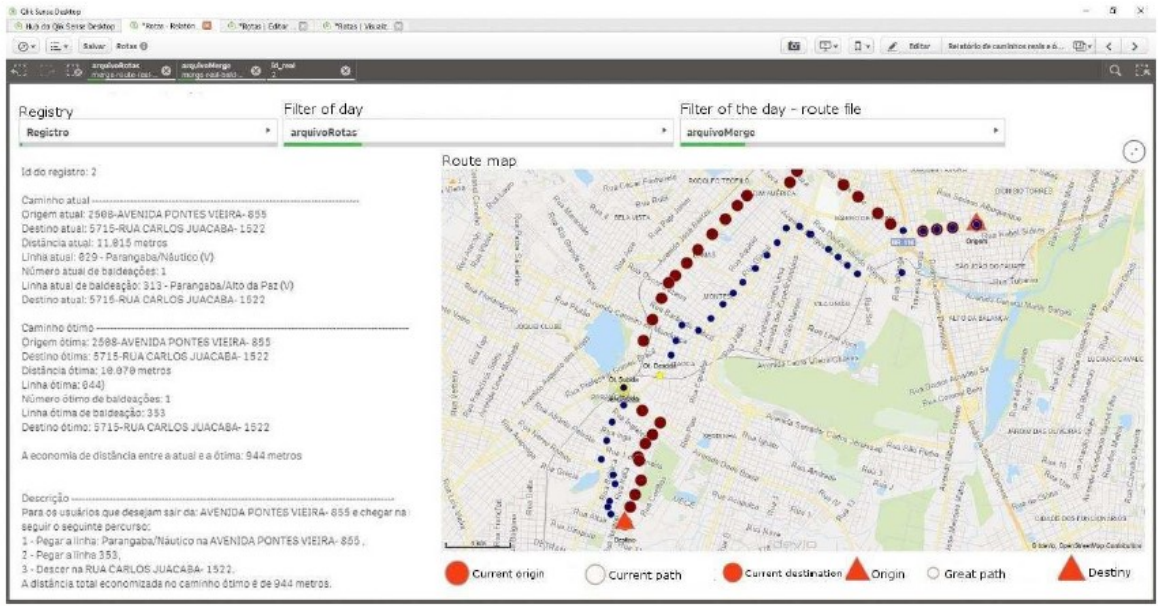

Figure 4: Report showing the actual route and optimal alternative

This report was important because researchers, when applying the questionnaire, would be able to clarify doubts about origin and destinations, ask about the riders' knowledge of optimal routes between that OD pair, and to show them the optional paths on a map. In addition to personal profile, such as education, age and sex, as well as information about familiarity with mobile technology, or any other useful information, such report would allow researchers to gather valuable information about what other environmental and/or local context characteristics that could explain why user take less than optimal routes.

**5. METHOD EFFECTIVENESS VALIDATION**

As an attempt to validate the benefits generated by our methodology to surveys' data gathering, this section aims to estimate the probability ($P(p)$) of finding riders who take routes considered unsatisfactory after leaving a bus stop $p$. We estimate the chance of meeting these riders in two ways, first randomly, and second thorough our methodology. We verified that, in general, users make an optimal number of transfers, but do not choose the best place to do so. This tendency leads to actual routes that are longer than the optimized ones, for this reason travel distances will serve us as criteria ($\lambda$) for evaluating actual ($C_r$) and optimal ($C_o$) routes. Moreover, we also consider as benefit of our methodology the drop in the cost for the application of survey questionnaires. We measure this cost indirectly, by the number of bus stops necessary for the surveys viability. It is logical to assume that the more bus stops surveyed, the higher the costs involved. Considering this, we simulated four scenarios varying the amount of bus stops, among 10, 100, 500, and 1000, for a survey application.

Figure 5 illustrates the estimation results. In all four graphs, the x-axis is the quality threshold $\lambda$ and the y-axis represents the probability *P(p)*. The green dots highlight the results of our methodology, and the red dots show the results achieved by a random selection of bus stops. Results indicate that the proposed methodology presents higher probabilities in all investigated scenarios. Furthermore, the lower the number of surveyed bus stops, the more our methodology stands out in comparison to a random methods.

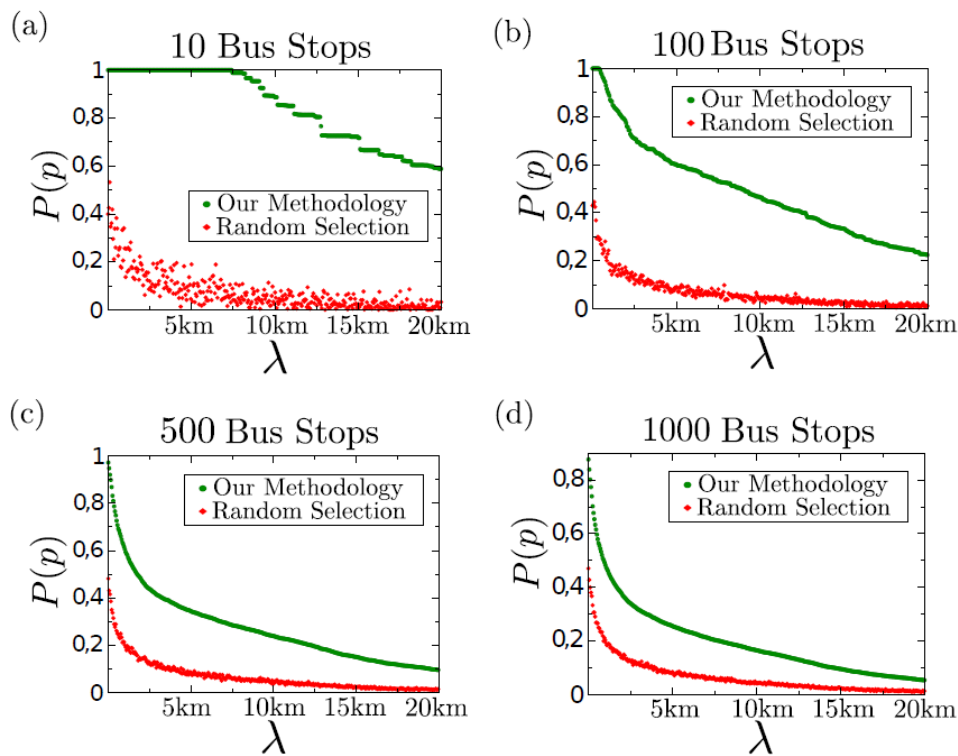

Figure 5: Simulation Scenarios.

It is clear that probabilities decrease as the $\lambda$ threshold, as well as, the number of visited bus stops increase. We can also say that the fewer the surveyed bus stops, the better the comparative results presented by our methodology. Fewer surveyed bus stops means lower survey costs. For the 10 stops simulation (Figure 5a), at a $\lambda$ of 10km, the probability of finding unsatisfied PT riders using our methodology is over 80%, whereas for random encounters, it is below 10%. For the 1000 stops simulation (Figure 5d) the probabilities are still advantageous when applying our methodology, even if at a smaller scale.

## 6. POST QUESTIONNAIRE RESULTS

In the attempt to exemplify the method, we applied a survey on five bus stops in the city of Fortaleza. The limited number of surveyed stops was a necessity so that we could verify any benefits in finding unsatisfied PT riders. We selected different times of the day to perform the surveys according to the times of use depicted by the analyzed data. In all visited stops, we interviewed 28 passengers, from which, 14 presented potentially unsatisfactory experiences. The fact that we could find this amount of people reinforces the method effectiveness, addressed in section 5. Due to the way we reconstructed actual routes (Ref.), focusing in recurrent riders (who followed the same itinerary every day), all the 14 potentially unsatisfied PT riders had extensive experience with their route, which invalidates the lack of information about the existence of a better available route as a valid hypothesis to explain the riders' sub-optimal behavior. Given the absence of clear information available at the stop, users declared that they found information by trial and error (own experience) or accessing mobile applications that gave information about alternative routes. The assumption that the lack of information is a problem when not being able to choose the optimal route seems to make more sense to inexperienced users (e.g. those who are making the journey for the first time). So that we may confirm this experiment's statistical validity, more extensive sampling and analysis are necessary.

In general, for Fortaleza, when a passenger chooses a route, an important preference criterion is the number of connections along the route, as indicated by quantitative data analysis. However, only qualitative analysis could show that this preference is not definitive or exclusive. For example, when asked whether they would be willing to swap to another stop in order to take a time-efficient route, users responded negatively. The main alleged reasons related to comfort issues. This notion of comfort comes from different perspectives. Some respondents argued that the probability of a bus that followed the optimal route being crowded on arrival was too high. This would make the bus stop at the non-optimal route more attractive, as they would be able to embark and sit throughout the ride. Other users indicated different comfort-related factors, such as the existence of benches and shade at the actual-route stops. In other words, both (comfort-related) environmental aspects and travel efficiency (in terms of one of the studied preference criteria) are important aspects that affect PT riders' route choices.

An unusual response, which opens up interesting possibilities when designing the service, was a passenger who said she chose a clearly non-optimal route because she wanted to accompany her friends. Thus, there is an aspect related to sociability, which is rarely taken into account by those who design this type of service. The answers provided by passengers shine a light on important aspects to be considered by PT operators and decision makers so that services are in line with the preferences of users in this context. Furthermore, the acquisition of such rich qualitative data should be useful in fine-tuning PT allocation models.

## 7. DISCUSSION AND CONCLUSION

Cities are increasingly adopting programs to become smarter. Assuming that a smart city provides services in an environment where people have pleasant experiences, finding ways to capture how these experiences are being experienced is essential. Doing this in a rich environment of user profiles and with large volumes of data, such as those found in large cities, can be expensive and time consuming. Any benefits on problem identification, data collection and cost efficiency are welcome. The proposed methodology, while supported by data visualization tools, and as verified in the presented application work as an auxiliary tool to help planners and researchers in the tasks of, identifying the location of riders potentially facing problems with the PT system, and constructing hypothesis to why those riders are having unsatisfactory experiences. Furthermore, another output is the help it provides in planning on-site survey with better probabilities of useful data collection at potentially lower costs. We observed evidence to support that the proposed methodology increases the likelihood for the researchers at on-site surveys to find PT riders with potentially more useful information for the PT system planner or decision maker. We believe that the discrepancies between actual and optimized routes (both defined by the methodology) may signify that specific characteristics of the environment and local contexts have an impact on riders behavior. By identifying these riders, researchers can become aware of inconveniences that influence the user's behavior. These declared preferences, while certainly enables researchers to characterize PT riders, they may also indicate (e.g.) deficiencies in the provided services or the need for new services and technologies, to make the experience more convenient.

A necessary next step is the large-scale application of the method, once we have already identified potentially unsatisfied PT riders at, what seems to be, problematic bus stops, with the application of the method. In the pilot study, a survey conducted at only five of these highlighted stops led to the discovery of environmental and systemic characteristics (pointed out by unsatisfied PT riders) responsible for the less-than-optimal behavior of PT riders, which had not previously been identified by planners and city administrators. The notion of comfort relating to the absence of overcrowding and the quality of bus stops are two examples of data collected from the conducted pilot survey, which helped us identify, in a quicker and cheaper way, environmental aspects (in the city's supply network and transport infrastructure) that played important roles in PT riders' route choices.

As an extension to this research, we see in this methodology a great potential for the development of multi-criteria route choice models. From the identification of discrepancies between actual and optimal routes, associated with the results of large-scale on-site surveys, we believe it to be possible for modelers to readjust the route-optimization algorithms with the environmental and local-context characteristics described by the unsatisfied PT riders. This readjust could happen as penalties applied to PT network elements such as bus stops or certain links, applicable to certain categories of users (agents) during the optimization process. This way, not only we could acquire a potentially precise rout-choice modeling tool, it could also mean, when paired with the polygon intersection tool, described in section 3.4, a useful validation tool.

We believe that the contribution offered by this work aims the scientific community interested in studying the quality of services, and that make use of ethnographic practices. This work allows managers to look beyond the set analysis, allowing them to make critical readings of social contexts and of the use of the service under study. In fact, the cyclical nature of the methodology provides the possibility of reanalysis of collected data with variations in criteria and assumptions. The cyclical application of the methodology will be part of further work still in the context of urban mobility. In the long term, the application of the methodology in other contexts serves to validate its reach and applicability.

**ACKNOWLEDGEMENT**

This research would not be possible without the full support offered by UNIFOR.**Bibliography**

1. Brands T, de Romph E, Veitch T, Cook J. Modelling Public Transport Route Choice, with Multiple Access and Egress Modes. *Transp Res Procedia*. 2014;1(1):12-23. doi:10.1016/j.trpro.2014.07.003.

2. Cascetta E. *Transportation Systems Analysis: Models and Applications*. Vol 29. 2nd ed. Springer; 2009.

3. Tyrinopoulos Y, Antoniou C. Public transit user satisfaction: Variability and policy implications. *Transp Policy*. 2008;15(4):260-272. doi:10.1016/j.tranpol.2008.06.002.

4. Nyarirangwe M, Mbara T. Public Transport Service Modal Choice , Affordability and Perceptions in an Unpalatable Economic Environment : the Case of an Urban